\documentstyle[aaspp4,11pt]{article}

\def\HI{\protect\ion{H}{1}}

\def\NaI{\protect\ion{Na}{1}}
\def\MgII{\protect\ion{Mg}{2}}

\def\PII{\protect\ion{P}{2}}
\def\ClI{\protect\ion{Cl}{1}}
\def\ClII{\protect\ion{Cl}{2}}
\def\CaII{\protect\ion{Ca}{2}}
\def\TiII{\protect\ion{Ti}{2}}
\def\MnII{\protect\ion{Mn}{2}}
\def\FeII{\protect\ion{Fe}{2}}
\def\FeIII{\protect\ion{Fe}{3}}
\def\dex#1{10$^{#1}$}
\def\tdex#1{$\times$10$^{#1}$}

\def\cmm#1{cm$^{-#1}$}
\def\kms{km\,s$^{-1}$}

\def\navg{$\langle n_{\rm H}\rangle$}

\begin{document}


\title{Dependence of Gas Phase Abundances in the ISM on Column Density}

\author{B.P. Wakker, J.S. Mathis}
       \affil{Department of Astronomy, University of Wisconsin \\
       475 N Charter St, Madison, WI\,53706, USA \\
       wakker@astro.wisc.edu, mathis@astro.wisc.edu}

\begin{abstract}
Sightlines through high- and intermediate-velocity clouds allow measurements of
ionic gas phase abundances, $A$, at very low values of \HI\ column density,
N(\HI). Present observations cover over 4 orders of magnitude in N(\HI).
Remarkably, for several ions we find that the $A$ vs N(\HI) relation is the same
at high and low column density and that the abundances have a relatively low
dispersion (factors of 2--3) at any particular N(\HI). Halo gas tends to have
slightly higher values of $A$ than disk gas at the same N(\HI), suggesting that
part of the dispersion may be attributed to the environment. We note that the
dispersion is largest for \NaI; using \NaI\ as a predictor of N(\HI) can lead to
large errors.
\par Important implications of the low dispersions regarding the physical nature
of the ISM are: ($a$) because of clumping, over sufficiently long pathlengths
N(\HI) is a reasonable measure of the {\it local} density of {\it most} of the H
atoms along the sight line; ($b$) the destruction of grains does not mainly take
place in catastrophic events such as strong shocks, but is a continuous function
of the mean density; ($c$) the cycling of the ions becoming attached to grains
and being detached must be rapid, and the two rates must be roughly equal under
a wide variety of conditions; ($d$) in gas that has a low average density the
attachment should occur within denser concentrations.
\end{abstract}



\section{Introduction}
The study of elemental abundances in the interstellar medium has a long history,
both in the optical and with space-based UV spectrographs, such as {\it
Copernicus} and the Hubble Space Telescope ({\it HST}). These studies show that
most elements have apparent abundances relative to hydrogen substantially below
those in the Sun. This is interpreted as due to depletion of ions onto dust
grains.
\par Savage \& Sembach (1996a) summarized high-quality abundance studies based
on {\it HST} data. They concluded that the depletion of an element depends on
the nature of the interstellar medium along the sightline. In cold, dense disk
gas, S has near-solar abundance, while Si, Mg, Mn, Cr, Fe and Ni are
progressively more depleted. In warm-disk gas, the gas-phase abundances appear
higher than in the cool gas, while in halo gas they are even larger.
\par Jenkins, Savage \& Spitzer (1986) used {\it Copernicus} to measure accurate
column densities for \MgII, \PII, \ClI, \ClII, \MnII\ and \FeII, for stars with
distances, $d$, between 0.1 and 2\,kpc. They measured N(\HI) from the damping
wings of Ly$\alpha$ and found a relation between the ionic abundance relative to
\HI, $A$, and the average density, \navg\ ($\equiv$ N(\HI)/$d$). An explanation
was provided by Spitzer (1985), who suggested that the observed depletion in a
line of sight depends on the fraction of warm and cold (dense) gas encountered,
where each component has a specific gas phase abundance of each ion.
\par Crinklaw, Federman, \& Joseph\ (1994) observed \CaII\ and \TiII\ toward 12
stars with distances between 150 and 650\,pc. The slope of the relation between
log[\navg] and log[$A$(ion)] is similar for \CaII\ and \TiII, but it differs
from those for \FeII, \MnII, \MgII\ and \PII. They proposed that \CaII\ and
\TiII\ mostly occur in the warm intercloud medium, and that the apparent
depletion is very sensitive to the inclusion of high-density regions (with high
depletion) in the sightline.
\par These studies were limited to sightlines with N(\HI) in the range \dex{20}
to 7\tdex{21}\,\cmm2. To extend this range, we used the high- and
intermediate-velocity gas (HVCs and IVCs, see Kuntz \& Danly 1996, Wakker \& van
Woerden 1997, Wakker 2000). In 21-cm emission these stand out in velocity from
gas in the Milky Way disk, so that weak components (down to \dex{18}\,\cmm2) can
be measured. At low velocities, parcels of gas with such low column densities
can not be separated from other high column density concentrations in the line
of sight.
\par Wakker (2000) analyzed all the published absorption-line data pertaining to
the high- and intermediate-velocity gas. Improved \HI\ column densities (and
thus improved ion abundances) were determined for about 250 sightlines. \CaII\
was measured for 97 components, \TiII\ for 6, \FeII\ for 13, \MnII\ for 5,
\MgII\ for 10, and \NaI\ for 34 (low-velocity \NaI\ was taken from Ferlet et
al.\ 1985). Most IVCs have intrinsically near-solar abundance (Wakker 2000).
However, HVC complex~C has $A$$\sim$0.1 solar (Wakker et al.\ 1999), and the
Magellanic Stream has $A$$\sim$0.25 solar (Gibson et al.\ 2000). Including these
HVC complexes could lower some of the apparent abundances (in 8 cases for \CaII,
4 for \FeII\ and 10 for \MgII), but no clear effect is visible. [An anomalously
low value for \MgII, in the VHVC probed by Mrk\,205, was excluded as this cloud
may have low intrinsic abundance, and N(\HI) is uncertain.]
\par No distances are known for many of the HVCs/IVCs, and they fill only a
small fraction of the pathlength. Thus, we looked at the relation between $A$
and N(\HI), rather than \navg. Unexpectedly, we found tight anticorrelations of
N(\HI) with the abundances of \MgII, \CaII, \TiII, \MnII\ and \FeII. A minimal
correlation is expected because both quantities are integrated along lines of
sight with varying physical conditions. Variations include ($a$) the ionization
of H, ($b$) widely differing histories of grain destruction by interstellar
shocks, and ($c$) considerable differences of the interstellar radiation field
impinging on the gas. \CaII\ and \NaI\ are not even the dominant stage of
ionization in neutral regions, so the radiation field is of critical importance
for their ionization fraction. In \S2 we show the correlations between column
density and abundances, while in \S3 we discuss some implications.


\section{Correlations between abundances and N(\HI)}
\subsection{The data}
Figure~1 shows scatter plots of log[N(\HI)] vs log[$A$(ion)], the gas-phase
abundance relative to \HI, for \MgII, \CaII, \TiII, \MnII, \FeII\ and \NaI. The
least-squares fit and dispersion (labeled ``rms'') are given in each panel.
Symbols indicate the origin of the datapoints, as detailed in the figure
caption. The Local ISM points (open squares) were not used in the fits (see
\S2.5 below). The high-quality data obtained with the ``Goddard High Resolution
Spectrograph'' ({\it GHRS}) are presented separately, in order to show the
difference between halo and disk sightlines more clearly (\S2.5).
\par The currently preferred oscillator strengths of \MnII-1197, 1201 and
\MgII-1239, 1240 are 0.20 and 0.67 dex higher than those used by Jenkins et al.\
(1986) (see Savage \& Sembach 1996a). We therefore corrected their \MnII\ and
\MgII\ column densities downward by these amounts.
\par As the highest abundances are associated with high-velocity gas, we might
be seeing a manifestation of the Routly \& Spitzer (1952) effect (this is the
increase in the ratio N(\CaII)/N(\NaI) with LSR velocity). This effect is
interpreted as showing that Ca is less depleted at higher peculiar velocities.
However, it is based on nearby ($<$100\,pc), low-velocity ($<$20\,\kms) gas
(e.g.\ Vallerga et al.\ 1993). For the HVCs/IVCs the LSR velocity is not a good
measure of the peculiar velocity relative to their surroundings. Further, the
IVCs for which both \NaI\ and \CaII\ have been measured do not show the
effect (Wakker 2000).
\par We now discuss some implications of the $A$ vs N(\HI) relations.

\subsection{Result 1: small dispersions}
For \MgII, \MnII\ and \FeII\ the standard deviations of log[$A$(ion)] from the
least-squares fit at a given N(\HI) are only 0.27 in the log. That is less than
a factor 2 either way, even though the correlation extends over 4 orders of
magnitude in N(\HI) and over 2 orders of magnitude in $A$(\FeII). The standard
deviations are slightly larger for \CaII\ and \TiII\ (0.4 in the log, or a
factor 2.6 either way). The largest scatter is seen for \NaI, where it is 0.52
in the log or a factor 3.5 either way.
\par The larger scatter for \CaII\ and especially \NaI\ may be related to the
fact that, unlike the other ions, these are not the dominant ionization stage in
the diffuse ISM, so that ionization effects may play a larger role. However,
\TiII\ is the dominant ionization stage, and it shows a scatter that is
comparable to that of \CaII.
\par Ferlet et al.\ (1985) proposed a slope near zero for the relation between
\NaI\ abundance and N(\HI) for the low-velocity gas. This has been used
extensively to estimate N(\HI) from N(\NaI). The lower column density IVC data
tend to show slightly higher abundances. However, the spread in $A$(\NaI) at any
given value of N(\HI) is rather large. For a given N(\NaI), the derived N(\HI)
will be a factor $>$3.5 off either way in 33\% of the cases. The most deviant
points in the \NaI\ diagram differ from the mean relation by a factor $>$20.
Thus, although on average N(\NaI)/N(\HI) is independent of N(\HI), N(\NaI) is a
rather poor predictor of N(\HI).

\subsection{Result 2: differing slopes}
The slopes of the correlations differ substantially: $-$0.78$\pm$0.04 for \CaII,
$-$0.69$\pm$0.08 for \TiII, $-$0.59$\pm$0.04 for \FeII, $-$0.39$\pm$0.04 for
\MnII, $-$0.24$\pm$0.04 for \MgII, and $-$0.16$\pm$0.06 for \NaI. We find slopes
of $-$0.63$\pm$0.12 and $-$0.66$\pm$0.23 for low- and high-velocity \CaII; for
\FeII\ we find $-$0.54$\pm$0.05 and $-$0.70$\pm$0.07, respectively. For other
elements there are insufficient datapoints to make separate fits. The errors are
larger because of the reduced range in N(\HI), but the resulting slopes are the
same to within the formal error. This shows that on average the high column
density, low-velocity disk gas behaves in the same manner as the low column
density high-velocity halo gas.

\subsection{Result 3: no obvious ionization effects}
The ionization potentials (I.P.) of \FeII, \MnII, and \MgII\ are 16.2, 15.6, and
15.0\,eV, respectively (compared to 13.6\,eV for \HI), so these ions can
co-exist with H$^+$. The \FeIII/\FeII\ ratio should not depend on N(\HI), but
only on the surrounding radiation field (which only depends on location). Thus,
if the interstellar radiation field has similar values in the sightlines that
were observed, and if H$^+$ were only present in shells around \HI\ cores, then
the fraction of H$^+$ should increase with decreasing N(\HI), resulting in an
overestimate of the abundance, when calculated as N(ion)/N(\HI). This would
produce an upturn at low N(\HI) in the $A$ vs N(\HI) relation. However, such an
effect is not obvious. \CaII\ (I.P.\ 11.9\,eV) should ionize more easily than H,
leading to a downturn at low N(\HI). This is also not seen.

\subsection{Result 4: environmental dependence}
For \MgII, \MnII\ and \FeII\ the {\it GHRS} disk points (open circles in Fig.~1)
scatter around those of Jenkins et al.\ (1986). {\it GHRS} halo sightlines
(filled circles) tend to lie above the average correlation. The small scatter in
$A$(\FeII) remarked upon by Savage \& Sembach (1996a) is consistent with the
fact that their halo points span only a factor 10 in N(\HI). However, some more
recent halo points (filled triangles) have higher \FeII/\HI\ ratios, as do some
HVCs.
\par The low N(\HI) Local ISM (LISM; open squares) shows abundances far below
halo points at the same N(\HI), although they are near the high end for disk
gas and the observed depletion is consistent with the average Local ISM density
of 0.1\,\cmm3. We surmise that both the Sun and the LISM probes lie within a
cloud, so that the LISM paths do not pass through all of it, unlike the case for
clouds in the sightlines to more distant stars. Thus, the Local ISM points
suggest that density is the relevant physical parameter determining depletion.
Over sufficiently long pathlengths column density can substitute because it is
dominated by the densest regions.
\par The largest depletions are seen toward $\zeta$ Oph and $\xi$ Per, which
were used by Sembach \& Savage (1996) as the archetypical cold disk sightlines.
These have high N(\HI) and $A$(\MgII) and $A$(\FeII) lie below the average
relation. Including N(H$_2$) would move these points 0.2 and 0.4 dex to the
right and down, respectively, shifting them $\sim$0.1 and 0.2 dex further below
the average.
\par We thus see the environmental dependence found by Sembach \& Savage (1996):
disk sightlines tend to have low gas-phase abundances, while halo sightlines
show high gas-phase abundances. We can understand this in large part as a column
density dependence, as disk sightlines tend to have higher column density than
halo sightlines.


\section{Discussion}
The general trend that the gas phase abundances of refractory ions decreases
with N(\HI) is, of course, well known; the ions are removed from the gas phase
by sticking to grains. {\it However, the low dispersions of the depletions at a
particular value of N(\HI), continuing to low values to N(\HI), are remarkable.}
The depletion patterns provide a diagnostic of the density fluctuation within
the ISM, since the depletions require encounters between the grains and ions.
{\it The low dispersions of $A$ at a given N(\HI) suggest additional
properties of the ISM:}
\par 1. Large values of N(\HI) arise from one or a few regions of relatively
large local density rather than from a superposition of many regions of lower
density. In this way, there is unexpected physical meaning to the integrated
column density: a good correlation with the mean {\it local} density of the
typical H atom along the sightline. If large values of N(\HI) were built up from
many regions of low density along the line of sight, the averaged $A$ would
correspond to the lower densities. The association of large N(\HI) with a large
local density is strengthened by the fact that the anticorrelation of gas phase
abundances with \navg\ is better than with N(\HI) (Savage \& Bohlin 1979).
\par 2. The low dispersions in log[$A$] suggest that grains are not destroyed in
strong shocks, as some models have suggested. If such destruction occurred, a
few points with large gas phase abundances would be expected at large N(\HI).
Rather, the detachment of a refractory element from grains seems to be a
continuous function of the mean density, and is common enough to produce a low
dispersion about the mean. Grain destruction may occur by means of many small
shocks with a broad distribution of velocities.
\par 3. Both the rates of ions sticking to grains and becoming detached from
them (probably in atomic form) must be rapid enough to achieve rather good
statistical equilibrium. These balancing rates lead to a particular value of $A$
for a given density; otherwise, the past history of the \HI\ along the line of
sight would strongly influence $A$.
\par 4. In view of (3) above, we think that the best possibility is that
attachment takes place within regions of far greater than average density
distributed along each line of sight. Encounters between grains and ions are
very slow if the mean density of the diffuse ISM is assumed (see below).
Possibly magnetohydrodynamic waves create non-thermal motions of the ions
relative to the grains, since the charge per mass of the grains is vastly less
that for the ions. As a result, ions encounter more dust grains and the net
attachment rate increases.
\par Under the assumption that relative motions were thermal, Weingartner \&
Draine (1999) calculated average gas phase abundances of a highly depleted
species (\TiII) for large N(\HI). In their model, dense
(\navg$\sim$\dex3\,\cmm3) molecular clouds contain half the mass of the ISM,
30\% is in cool, neutral clouds (\navg=30\,\cmm3) and 20\% is in warm neutral
clouds (\navg=0.4\,\cmm3). The attachment of the Ti onto grains occurs in the
molecular and cool clouds. This model shows that the  observed $A$(\TiII) for
large N(\HI) can only be achieved by circulation into very dense regions. To
achieve rapid attachment within the diffuse ISM, such circulation seems
necessary, operating at both low and high column densities. An additional
condition for rapid attachment of ions is that the gas be cold as well as dense,
in order to provide efficient focusing by the Coulomb force onto negatively
charged grains.

\acknowledgments We appreciate comments and clarifications by an anonymous
referee, Alex Lazarian, Chris Howk, and Bruce Draine. B.P.W.\ is partially
supported by NASA grant NAG5-9179.




\newpage
{Figure 1. 
Correlation plots of the abundances of \NaI, \MgII, \CaII, \TiII, \MnII\ and
\FeII\ vs N(\HI). The solid lines show the least-squares fits, whose
coefficients are given by the label in the lower left corner. ``rms'' is the rms
of the residual, also indicated by the two parallel dotted lines. ``$\rho$'' is
the correlation coefficient. Filled circles are for the HVC and IVC data from
Wakker (2000). If measurement errors were given for N(ion), error bars are
shown; usually no errors are available. Crosses show data for low-velocity gas
from Ferlet et al.\ (1985; \NaI), Jenkins et al.\ (1986; \MgII, \MnII\ and
\FeII) and Crinklaw et al.\ (1994; \CaII, \TiII). Errors are usually smaller
than the size of the dot. The right panels for \MgII, \MnII\ and \FeII\
separately show measurements with relatively low systematic errors obtained with
the {\it GHRS} by Savage et al.\ (1993), Sofia et al.\ (1994), Cardelli et al.\
(1995), Spitzer \& Fitzpatrick (1995), Savage \& Sembach (1996b) and Sembach \&
Savage (1996). For these data, halo sightlines are shown by filled circles, disk
sightlines by open circles. Filled/open triangles show halo/disk sightlines from
Fitzpatrick \& Spitzer (1997; after combining their fit components into the 5
\HI\ components visible in the spectrum) and Howk et al.\ (1999). Open squares
show Local ISM sightlines from Linsky et al.\ (1995), H\'ebrard et al.\ (1999)
and Howk (priv.\ comm.); these were not used in the fits, see \S2.4. For all
low-velocity data the error bars are smaller than the symbol sizes. The
horizontal lines show the solar reference abundance, from Anders \& Grevesse
(1989).}
\newpage\vbox to \vsize{\vss{Figure 1}\includegraphics{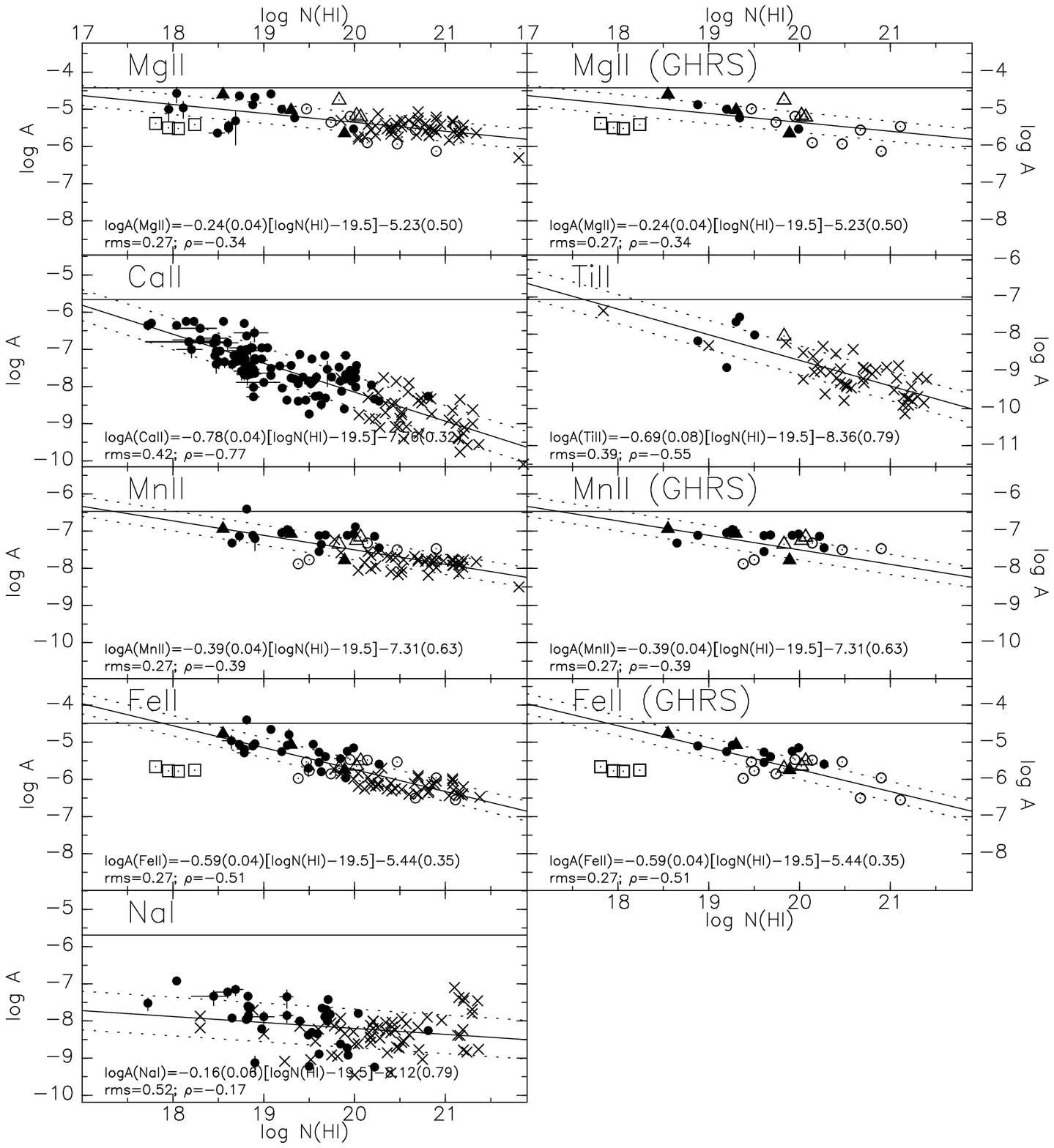}} 


\begin{thebibliography}{}
\bibitem[x (2000)]{x}Anders, E., Grevesse, N.\
                     1989, Geochim.\ Cosmochim.\ Acta, 53, 197
\bibitem[x (2000)]{x}Cardelli, J.A., Sembach, K.R., \& Savage, B.D.\
                     1995, \apj, 440, 241
\bibitem[x (2000)]{x}Crinklaw, G., Federman, S.R., \& Joseph, C.L.\
                     1994, \apj, 424, 748
\bibitem[x (2000)]{x}Ferlet, R., Vidal-Madjar, A., \& Gry, C.\
                     1985, ApJ, 298, 838
\bibitem[x (2000)]{x}Fitzpatrick, E.L., \& Spitzer, L.\ 1997, \apj, 475, 623
\bibitem[x (2000)]{x}Gibson, B.K., Giroux, M.L., Penton, S.V., Putman, M.E., Stocke, J.T., Shull, M.J.\
                     2000a, AJ, accepted, astro-ph/0007078
\bibitem[x (2000)]{x}H\'ebrard, G., Mallouris, C., Ferlet, R., Koester, D., Lemoine, M., Vidal-Madjar, A., York, D.\
                     1999, \aap, 350, 643
\bibitem[x (2000)]{x}Howk, J.C., Savage, B.D., \& Fabian, D.\
                     1999, \apj, 525, 253
\bibitem[x (2000)]{x}Kuntz, K.D., \& Danly, L.\ 1996, \apj, 457, 703
\bibitem[x (2000)]{x}Jenkins, E.B., Savage, B.D., \& Spitzer, L.\
                     1986, \apj, 301, 355
\bibitem[x (2000)]{x}Linsky, J.L., Diplas, A., Wood, B.E., Brown, A., Ayres, T., Savage, B.D.\
                     1995, \apj, 451, 335
\bibitem[x (2000)]{x}Routly, P.M., Spitzer, L.\
                     1952, \apj, 115, 227
\bibitem[x (2000)]{x}Savage, B.D., Lu, L., Weymann, R., Morris, S., \& Gilliland, R.\
                     1993, \apj, 404, 124
\bibitem[x (2000)]{x}Savage, B.D., \& Bohlin, R.C.\ 1979, \apj, 229, 136
\bibitem[x (2000)]{x}Savage, B.D., \& Sembach, K.R.\ 1996a, ARAA, 34, 279
\bibitem[x (2000)]{x}Savage, B.D., \& Sembach, K.R.\ 1996b \apj, 470, 893
\bibitem[x (2000)]{x}Sembach, K.,R., \& Savage, B.D.\ 1996, \apj, 457, 211
\bibitem[x (2000)]{x}Sofia, U.J., Cardelli, J.A., \& Savage, B.D.\
                     1994, \apj, 430, 650
\bibitem[x (2000)]{x}Spitzer, L.\ 1985, \apj, 290, L21
\bibitem[x (2000)]{x}Spitzer, L., \& Fitzpatrick E.L.\
                     1995, \apj, 445, 196
\bibitem[x (2000)]{x}Vallerga, J.V., Vedder, P.W., Craig, N., Welsh, B.Y.\
                     1993, \apj, 411, 729
\bibitem[x (2000)]{x}Wakker, B.P., Howk, C., Savage, B.D., Tufte, S.L.,
                     Reynolds, R.J., van Woerden, H., Schwarz, U.J., Peletier,
                     R.F., Kalberla, P.M.W.\ 1999, Nature, 400, 388
\bibitem[x (2000)]{x}Wakker, B.P., \& van Woerden, H.\ 1997, ARAA, 5, 217
\bibitem[x (2000)]{x}Wakker, B.P.\ 2000, ApJS, submitted
\bibitem[x (2000)]{x}Weingartner, J.C., \& Draine, B.T.\ 1999, \apj, 517, 292
\end{thebibliography}
\end{document}